\documentclass[12pt]{article}
\usepackage[utf8]{inputenc}
\usepackage[english]{babel}
\usepackage{amsfonts}
\usepackage{amssymb,enumitem}
\usepackage{amsmath,amsthm}
\usepackage{booktabs}
\usepackage{graphicx}
\usepackage{titlesec}
\usepackage{adjustbox}

\usepackage{multirow}
\usepackage{caption}
\usepackage{cite}

\usepackage{amsthm,amsmath}

\usepackage[utf8]{inputenc} 

\usepackage{tikz}
\usepackage{graphicx,subfigure}
\usepackage{float}
\usepackage{multirow}
\usetikzlibrary{arrows}
\title{A Hybrid Compartmental Model with a Case Study of COVID-19 in Great Britain and Israel}
\author{ Greta Malaspina\footnote{Department of Mathematics and Informatics, Faculty of Sciences, University of Novi Sad, Novi Sad, Serbia. Email: greta.malaspina@dmi.uns.ac.rs. }, Stevo Rackovi\'c\footnote{Institute for Systems and Robotics, Instituto Superior Técnico, Lisbon, Portugal. Email: stevo.rackovic@tecnico.ulisboa.pt }, Filipa Valdeira\footnote{Department of Environmental Science and Policy, Università degli Studi di Milano, Milan, Italy. Email: filipa.marreiros@unimi.it }}
\date{}

\begin{document}
\maketitle

\begin{abstract} 
Given the severe impact of COVID-19 on several societal levels, it is of crucial importance to model the impact of restriction measures on the pandemic evolution, so that governments are able to take informed decisions.
Even though there have been countless attempts to propose diverse models since the raise of the outbreak, the increase in data availability and start of vaccination campaigns calls for updated models and studies. Furthermore, most of the works are focused on a very particular place or application and we strive to attain a more general model, resorting to data from different countries. In particular, we compare Great Britain and Israel, two highly different scenarios in terms of vaccination plans and social structure.
We build a network-based model, complex enough to model different scenarios of government-mandated restrictions, but generic enough to be applied to any population. To ease the computational load we propose a decomposition strategy for our model.
\end{abstract}

\textbf{Keywords: COVID-19, SIR, Social Networks}
\section{Introduction}
\label{sec:intro}
When facing a severe pandemic, as we have seen recently with COVID-19, it is fundamental for the governments to take actions to prevent or contain its spread, while taking into account their impact from the socio-economic point of view. In this context, modelling the evolution of the spreading and the impact of different kinds of containment measures becomes an indispensable tool. The development of vaccines and early start of vaccination plans is also a crucial factor on the evolution of the pandemic and it was not available in the initial studies, calling for new models that may lead to different conclusions regarding what could be the best course of action.

Mathematical models where one can conduct simulations considering different scenarios regarding implemented measures are therefore of the utmost importance in order to make informed decisions. Consequently, since the beginning of the pandemic, there has been a vast number of papers and simulations concerning specifically the outbreak of COVID-19 (see, for instance, review paper \cite{ReviewPaperGeneral} and references therein). The approaches differ both in regard to properties of the population taken into account --- age groups, density, size and connectivity --- but also in the considered government imposed restrictions --- lockdown, closing of schools and workplaces or restrictions in public transport, mandatory wearing of masks and, finally, medical procedures, including mass vaccination. Among the most common approaches are compartmental \cite{SIR} and network based models \cite{networks}.

\paragraph*{Compartmental models}
In compartmental models, the population is partitioned into epidemiologically relevant classes. Then, a system of differential equations regulates how the cardinalities of such classes change over time and how they influence one another. A typical model in this class is Kermack-McKendrick SIR model \cite{SIR}, given by the following equations
\begin{equation}\label{eq:SIR}
\begin{cases}
\frac{d}{dt}S(t) = -\frac\beta N S(t)I(t)\\
\frac{d}{dt}I(t) = \frac\beta N S(t)I(t) - \gamma I(t)\\
\frac{d}{dt}R(t) = \gamma I(t),
\end{cases}
\end{equation}
where $S(t)$, $I(t)$, and $ R(t)$ represent the number of susceptible (S), infected (I), and recovered (R) individuals at time $t$, while parameters $\beta$ and $\gamma$ denote the infection and recovery rate, respectively.\\
Methods of this kind are of great interest because their simple structure makes them suitable for wide set of problems, and their deterministic nature allows some theoretical considerations regarding the asymptotic behavior of the number of infected individuals, as well as the existence of equilibrium points of the system. Furthermore, they can be easily modified to introduce additional classes or more complex equations, in order to get a more comprehensive description of the situation. For the COVID-19 outbreak, several extensions of this model have been proposed in literature, taking into consideration, for instance, the presence of a latency period of the virus \cite{latent}, the adoption of masks \cite{EikenberryInfectious} and vaccinations campaigns \cite{vaccines}, or restrictive measures and human behavior \cite{nonlinear}. \\ 

However, these methods usually rely on several assumptions regarding the population under study and the disease itself. Typical hypotheses are that the population is large and closed (i.e. there are no external influences on the system) and that the natural birth and death rates can be disregarded (e.g. because the considered time interval is short enough). Moreover the coefficients involved in the equations, e.g. $\beta$ and $\gamma$ in \eqref{eq:SIR}, are usually assumed to be deterministic, and the individuals are assumed to be uniformly distributed in space, so that in a given time interval each individual has an equal probability of meeting any other. Overall, to provide an accurate estimation, these models require good knowledge of the disease-related parameters and of the implications of restrictive measures, such as social distancing and travel limitations, over the average number of contacts between individuals. In practice, though, great variability has been observed in the transmission rate and other important parameters of COVID-19 \cite{variability}, so it may be difficult to estimate them accurately. At the same time, the inherent complexity of interactions between individuals and the large variety of restrictive measures enforced by governments at different times makes the assumption about uniform distribution of the population impractical.\\ 

\paragraph*{Network-based models}
To overcome this last issue, several network-based models \cite{ReviewPaperAgent} have been developed. In this setting, the population and the direct interactions between individuals are represented by graph, where the infection spread. The nature of the equations behind these methods is usually analogous to those behind compartmental models. The fundamental difference is that at each time step, the number of contacts between individuals of different classes does not simply depend on the size of said classes (e.g. the term $S(t)I(t)/N$ in \eqref{eq:SIR}) but also on the connections between nodes and the states that neighboring nodes occupy. The use of a network to model direct interactions between individuals provides a significantly more precise estimate of the number of contacts at each time-step, which can lead to better predictions of the spread of the infection even when accurate estimations of the disease-related parameters are not available. On the other hand, as these approaches require working with networks with as many nodes as the number of individuals in the population, they can become prohibitively expensive when considering large populations such as large regions or entire countries. Therefore, they are usually employed to model the infection at city or community level \cite{networks,schools}.

\paragraph*{Our approach}

The model that we consider in this paper comes from an extended SIR network-based approach. Regarding the compartmental component, besides the three standard states ($S, I$ and $R$) in which a node can be at a given time, we include several additional states and inter-states that take into account factors like incubation period of the virus and quarantine periods. These are crucial in the evolution of COVID-19. We also use different parameters for individuals belonging to different states, such as different susceptibility to symptoms and different recovery and death rate based on the current state. The parameters describing the virus behaviour are chosen within the range proposed in the literature and according to the provided data.

The population is approximated with a set of of graphs where each graph corresponds to a densely populated area (city) and each node represents an individual person\footnote{Note that, since we take only a fraction of population, in our simulation, one node actually stands for more than one person.}. The connections between nodes change over time according to the imposed governmental restrictions. Moreover, a SIR-like approach (analogous to the first equation of (1), with small $\beta$) is used to model the interaction between different graphs. This can be interpreted as having the population partitioned into densely populated cities and assuming that the number of interactions between individuals belonging to different cities is considerably smaller than the number of interactions between individuals coming from the same city.

Furthermore, we include in our model data from imposed measures, allowing us to include vaccination programs, contact tracing and testing, as well as movement restrictions. The latter are modeled by carefully selecting the networks topology, while the former are introduced by altering transitions and states of the existing models.

When considering large populations (e.g. entire countries) there is usually a trade-off between accuracy and computational time: running a model with larger population size (that is, closer to the actual size of the country) gives more accurate results but increases the computational cost. As the use of the graphs allows us to get a more accurate model of the local contacts between individuals and the strategy used to estimate the number of contacts between individuals in different cities is inexpensive, this partitioned approach helps us to increase the population size and improve the accuracy while keeping the computational load moderate.

\subsection{Related work}

\paragraph*{Model}
 Common extensions of SIR model are SEIR \cite{SEIR}, with an additional exposed ($E$) state, accounting for the period of exposure or virus incubation before developing the disease, and SEIRD \cite{SEIRD}, with an additional $D$ state, accounting for dead individuals. However, We base our model on a further extension, SEIRS+ \cite{schools}, where hospitalization and quarantine are also considered. These states are crucial for the particular COVID-19 outbreak, as they allow us to compare the number of hospitalized people and to model the effects of quarantine mitigation measures.

\paragraph*{Networks}  One common approach for modeling the virus spread in literature is to use multiplex networks \cite{MultiplexSingapore,BostonModel}, where each overlapping network describes a type of contact (household, workplace). In these cases, each agent usually has associated a series of individual properties (location, sex, age) and each day must be simulated in detail, by having the agent at each specific location and verifying the contacts in place. In this paper, the distinction is only made between children and adults, as the former will not get vaccinated, but this has no impact on the network. Besides, we just change the topology of a single network according to the combination of restrictions in place, avoiding the need to model each kind of specific contact and in this sense our work is closely related to  \cite{networks}. In \cite{networks} the authors propose a SEIR network-based model with four different networks representing the population depending on the restrictions enforced at a given time (in particular, they distinguish between the cases when no restrictions are in place, big events are forbidden, and social distancing or lockdown is enforced). In our model, we consider more than 4 different networks, thus allowing for better distinction of the measures in place. Furthermore, the underlying state model is SEIRS+, instead of SEIR.

\paragraph*{Vaccination}
Since the start of vaccination campaigns, several models have been proposed in literature that take into account the effect of vaccines on the spread of the virus \cite{vaccines, vacc1, vacc2, VaccinesNetwork}. Most of these works focus specifically on vaccinations, analysing how different vaccination rates impact the number of cases and studying the relationship between vaccine efficacy and number of administered vaccines, in order to determine if heard immunity can be achieved given a set of policies. In this paper, we keep into account the influence of vaccines on both the probability of contracting the virus and the probability that, once infected, person develops the symptoms. However, we prefer to keep a broader point of view, and not to focus solely on the effects of vaccination, taking into account the interaction between diverse measures.

\paragraph*{Transition parameters} The values of transition parameters between each model state are still not well defined for COVID-19. At the beginning of the pandemic there were several efforts to find transmission parameters, especially focusing on China (or more specifically Wuhan) \cite{ReadNovel,TangUpdated,kurt_immunity}. These initial estimates, however, have been continuously updated as detailed statistics became available for different countries and geographical regions. In general, contact rate is considered a decreasing function of time \cite{TangUpdated} since it is strongly influenced by the public awareness state, hence many papers propose fitting different values for different days of the pandemic \cite{vaccines}. It is also common to consider different parameters depending on the individual's characteristics (parameters calibrated by age are found to be an important factor on a good prediction \cite{ChangAustralia}). In  our case, the parameters are either constant for the entire timeline or data-dependent. We also compare our model for two different countries --- Israel and Great Britain --- in order to ensure a good generalization of the model. On the other hand, most studies focus on singular particular locations \cite{brazil,ontario,germany}, thus allowing for a finer parameter tuning --- however, one should keep in mind that such an approach can easily lead to overfitting.

\paragraph*{Population} In order to model the diversity of the population and contacts among individuals, some models resort to census data \cite{BostonModel,ChangAustralia,FergusonReport,BicherAustralia} and/or mobile location \cite{BostonModel}. In our case, we just require the percentage of adult population and the measures in place to predict the model evolution, thus leading to an easier adaptation of the model to different settings and populations. While we assume that some of the parameters are country-dependent and they still need to be fitted to the specific country data, we do not require any additional information, such as census data. 

\subsection{Contributions}

The contributions of this work are three-fold. In the first place, we propose a model that combines a typical compartmental model with a graph-based one. In this sense, we can go beyond a single city or region and mimic several densely populated areas with a relatively low level of connection. Since the interactions between separate areas are cheap to incorporate, increasing the number of regions (and so the simulation population) leads to only a linear growth of the computational cost. This allows the model to benefit from the higher accuracy given by modeling of the population with a network, while avoiding the high computational cost that usually derives by the employment of networks. When suitable data is available, this also allows the modeling of local (e.g. regional) policies, without major alterations to the model here presented.

A further contribution lies on the data used. We took into consideration two countries with significantly different structures and contrasting policies of containment, which have a great influence on the local spread of the disease. This allows us to strive for a more general model, staying away from overfitting. While we do have some country specific parameters, they all regard the population behavior, where it is reasonable to require different models. The virus related parameters are kept the same for both countries enforcing our confidence in their estimates.

Finally, we use the proposed model to investigate a number of hypothetical scenarios. We consider different governmental measures, such as levels of restrictions, vaccination rates and testing policies, and we analyze how these choices affect the evolution of the contagion, what are the joint effects when measures are combined and how they compare with each other.

\subsection{Outline}

This document is organized as follows. In Section~\ref{sec:data_model} we describe our model and the datasets. In Section~\ref{sec:NumSim} we describe the process of fitting the model parameters and compare our prediction with ground truth data. In Section~\ref{sec:Scenarios} we test the impact of several measures by setting up a variety of different hypothetical scenarios. Finally, in  Section~\ref{sec:Discussion} we discuss the behaviour of the produced model and obtained results.

\section{Data and Model}
\label{sec:data_model}

The proposed model consists of several connected graphs, where each node of a graph represents a single individual in the population, and the edges between nodes correspond to contacts between people. At every day of the simulation a network-based SEIRS+ model is applied to each graph. The influence of different graphs on each other is simulated in the manner of a typical differential compartmental model. This section describes in detail how the model was built and how the COVID-19 statistics data was incorporated to fit the model parameters.

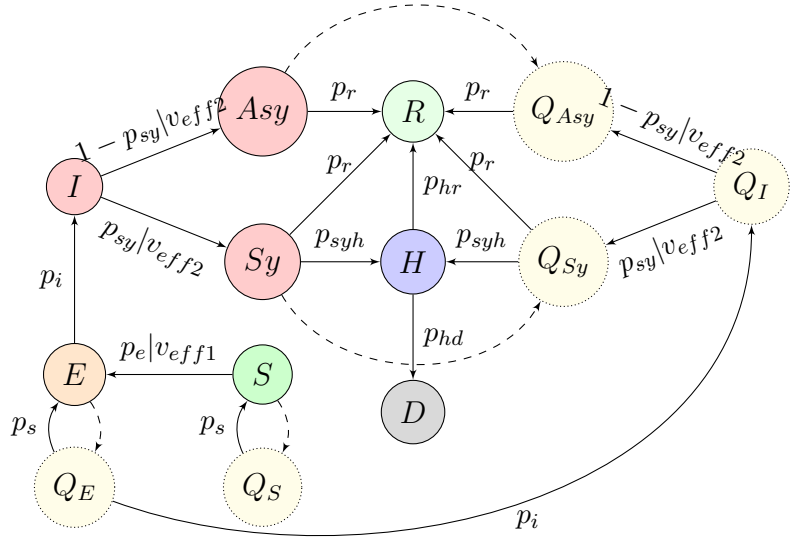
\begin{figure}
\centering
\begin{tikzpicture}
\tikzset{vertex/.style = {shape=circle,draw,minimum size=1.5em}}
\tikzset{edge/.style = {->,> = latex'}}
\node[vertex,fill=green!20] (S) at  (6,1.5) {$S$};
\node[vertex,densely dotted,fill=yellow!10] (Qs) at  (6,0) {$Q_S$};
\node[vertex,fill=orange!20] (E) at  (3.5,1.5) {$E$};
\node[vertex,densely dotted,fill=yellow!10] (Qe) at  (3.5,0) {$Q_E$};
\node[vertex,fill=red!20] (I) at  (3.5,4) {$I$};
\node[vertex,fill=red!20] (Asy) at  (6,5) {$Asy$};
\node[vertex,fill=red!20] (Sy) at  (6,3) {$Sy$};
\node[vertex,fill=green!10] (R) at  (8,5) {$R$};
\node[vertex,fill=blue!20] (H) at  (8,3) {$H$};
\node[vertex,fill=black!15] (D) at  (8,1) {$D$};
\node[vertex,densely dotted,fill=yellow!10] (Qsy) at  (10,3) {$Q_{Sy}$};
\node[vertex,densely dotted,fill=yellow!10] (Qasy) at  (10,5) {$Q_{Asy}$};
\node[vertex,densely dotted,fill=yellow!10] (Qi) at  (12.5,4) {$Q_{I}$};
\draw[edge] (S) to node [font=\small,above]{$p_e | v_{eff1}$} (E);
\draw[edge] (E) to node [font=\small,left]{$p_i$} (I);
\draw[edge] (Qs) to[bend left] node [font=\small,left]{$p_s$} (S);
\draw[edge] (Qe) to[bend left] node [font=\small,left]{$p_s$} (E);
\draw[edge,sloped, anchor=center] (I) to node [font=\small,above]{$1 - p_{sy} | v_{eff2}$} (Asy);
\draw[edge,sloped, anchor=center] (I) to node [font=\small,below]{$p_{sy} | v_{eff2}$} (Sy);
\draw[edge] (Asy) to node [font=\small,above]{$p_r$} (R);
\draw[edge] (Sy) to node [font=\small,above]{$p_r$} (R); 
\draw[edge] (Sy) to node [font=\small,above]{$p_{syh}$} (H);
\draw[edge] (H) to node [font=\small,right]{$p_{hr}$} (R);
\draw[edge] (H) to node [font=\small,right]{$p_{hd}$} (D);
\draw[edge] (Qasy) to node [font=\small,above]{$p_r$} (R);
\draw[edge] (Qsy) to node [font=\small,above]{$p_r$} (R);
\draw[edge] (Qsy) to node [font=\small,above]{$p_{syh}$}(H);
\draw[edge,sloped, anchor=center] (Qi) to node [font=\small,above]{$1-p_{sy} | v_{eff2}$} (Qasy);
\draw[edge,sloped, anchor=center] (Qi) to node [font=\small,below]{$p_{sy} | v_{eff2}$} (Qsy);
\draw[edge] (Qe) to[out=-20, in=-90] node [font=\small,below]{$p_i$} (Qi);
\draw[edge,dashed] (Sy) to[out=-60, in=-120] (Qsy);
\draw[edge,dashed] (S) to[bend left] (Qs);
\draw[edge,dashed] (E) to[bend left] (Qe);
\draw[edge,dashed] (Asy) to[out=60, in=120] (Qasy);
\end{tikzpicture}
\caption{Compartmental model}\label{fig:scheme}
\end{figure}

\subsection{Compartmental Model}
The design of our compartmental model is depicted in Figure \ref{fig:scheme}. Circles represent the possible states that each individual can occupy at any given time, and arrows represent possible transitions between states, with the parameters regulating such transitions. Dashed arrows indicate that a given transition between two states is not active during the whole simulation --- these are the transitions concerning the quarantine, hence we only activate them for the periods when the government policies demand so.
We consider the following states:
\begin{itemize}
\item $S$: susceptible  individuals --- people that have not been exposed to the virus but that could contract it upon contact with an infected person;
\item $E$: exposed individuals --- people that have been in contact with an infected individual and will become infected after the latency period of the virus has passed;
\item $I = Sy \cup Asy:$ infected individuals, divided into symptomatic and asymptomatic;
\item $H$: hospitalized individuals;
\item $R$: recovered individuals --- people that had the virus in the past, but that are now no longer infected nor susceptible;
\item $D$: deceased individuals;
\item $Q_S,\ Q_E,\ Q_{Asy},\ Q_{Sy}$: individuals that are susceptible, exposed, asymptomatic or symptomatic and that are currently in a state of quarantine.
\end{itemize}

For the sake of simplicity, let us first observe the scheme in Figure \ref{fig:scheme} under the setting of no imposed quarantine. For each individual in the population we have a Markov chain \cite{MarkovChains} with a source node $S$ and two sink nodes $R$ and $D$. Node $I$ is an intermediate state, and as soon as an individual jumps into $I$ it transitions to $Sy$ with probability $p_{sy}$ or to $Asy$ otherwise. To account for vaccination, we introduce two coefficients in our model: $v_{eff1}$ and $v_{eff2}$, appearing in some of the transition edges (see Section \ref{sss:vaccination}). Further, it is assumed that each individual can only get hospitalized if they previously manifested symptoms, and can only die if they were already hospitalized (that is, the only incoming edge to the $H$ state is from $Sy$, and the only incoming edge to $D$ is from $H$). If quarantine policies are in place, we see the analogous process but with $Q_{Asy}$ and $Q_{Sy}$. The following is a detailed list of the transition coefficients.
\begin{itemize}

\item $p_e$: probability that a susceptible person who has been in direct contact with an infected individual becomes exposed. We assume that this parameter changes over time, as to represent the variation in awareness of the population regarding the risks of transmission and the compliance of individuals with the restrictions (see Section \ref{sss:prob_exposure}). If the considered individual has been vaccinated, this probability is altered by the vaccine efficacy parameter $v_{eff1}$ (see Section \ref{sss:vaccination}).
\item $p_i$: probability that an exposed individual becomes infected --- this can be interpreted as the inverse of the expected latency period of the virus.
\item $ p_{sy}$: probability that an infected person develops symptoms. In case the considered individual is vaccinated, this probability is altered by the vaccine efficacy parameter $v_{eff2}$ (see Section \ref{sss:vaccination}).
\item $p_{syh}$: probability that a symptomatic individual gets hospitalized.
\item $p_r$: probability that an infected individual recovers --- this can be interpreted as the inverse of the average recovery time for people affected by the virus.
\item $p_{hr}$: probability that a hospitalized individual recovers --- this can be interpreted as the inverse of the average recovery time for hospitalized people.
\item $p_{hd}$: probability that a hospitalized individual dies.
\item $p_s$: probability that a quarantined person leaves the quarantine.
\end{itemize}

The transition between a class $X$ and its corresponding quarantined class $Q_X$ is modeled according to different possible testing policies, as well as contact tracing procedures that may be in place (see Section \ref{sss:testing_tracing}). 

\subsection{Datasets}\label{subsec:Datasets}
We used available data \cite{DataHub} for two countries, Great Britain (GBR) and Israel (ISR), over the course of more than a year. 

\begin{figure}[h]
    \centering
    \includegraphics[width=0.95\linewidth]{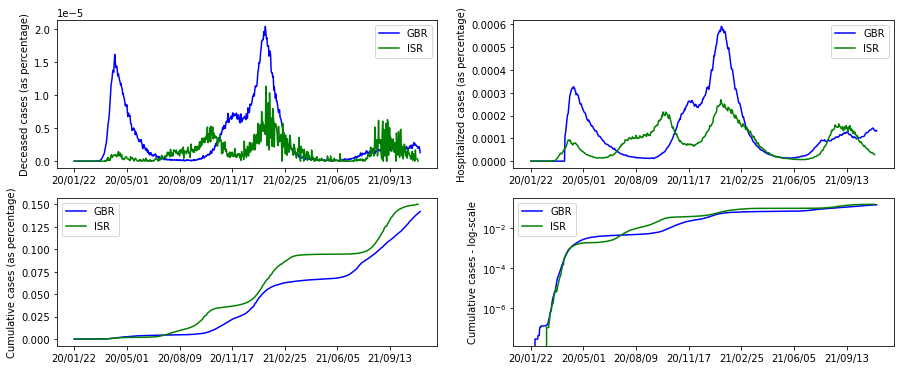}
    \caption{Ground truth attributes. In each subfigure $y-$axis represents values in terms of population percentage for each country, while $x-$axis provides the dates for the corresponding data. Upper left: new deceased cases at a given day. Upper right: hospitalized cases. Bottom left: cumulative count of confirmed cases. Bottom right: cumulative count of confirmed cases in a log-scale.}
    \label{fig:ground_truth}
\end{figure}

Since the goal is to study the effects of different policies on the evolution of the epidemic, we split the data attributes into two meaningful categories. The first group of attributes can be seen as potential inputs for the predictive model --- these are mainly governmental restrictions and policies in place, as well as the number of vaccinations and tests. The second group consists of attributes containing the number of infected, hospitalized, dead individuals, etc. These are potential ground truth (GT) values for the model. From this group, we mainly focus on the number of hospitalized and deceased individuals as GT. The choice lies on the fact that these are the most reliable values, since not all cases of the virus were actually detected, and there is no perfect way to trace recovered people. (We also do not account for particular cases of ICU (Intensive Care Unit) or people undergoing ventilation in our model because the counts are not consistent across the countries, so we do not make this distinction). Even though we do not directly use the count of confirmed cases in our model, we keep in mind that the number of predicted cases should not differ significantly from the available count. Figure \ref{fig:ground_truth} shows the GT values for both countries as a percentage of the total population. A cumulative number of cases is presented in the log scale to compensate for the exponential growth of the curve. For GBR we see three distinctive waves, with the second one having two peaks, while for ISR there are four. We further notice that both the numbers of hospitalized and deceased cases are lower for ISR than for GBR.

\begin{figure}[h]
    \centering
    \includegraphics[width=0.6\linewidth]{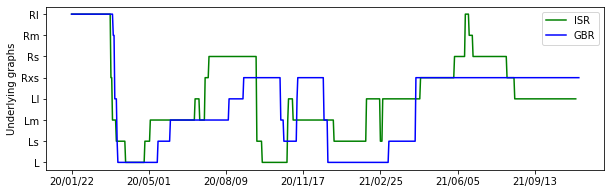}    
    \caption{First group of predictors. In each subfigure $y-$axis gives values in terms of population percentage for each country, and $x-$axis gives dates for the corresponding data.
    Left: number of performed tests for a day. Right: cumulative count of administered vaccines.}
    \label{fig:test_and_vaccines}
\end{figure}

As the predictors for our model we consider the following 3 groups of attributes: 
\begin{itemize}
    \item Vaccination and testing --- we have access to the number of administered vaccines and performed tests per day, as shown in Figure \ref{fig:test_and_vaccines}. Both countries started vaccination campaigns around the same day, but vaccination in ISR had a faster pace and added up to a higher overall number of doses compared to GBR. On the other hand, GBR had a higher and more consistent number of performed tests.
    \item Testing and contact tracing policies --- these are two ordinal attributes and Figure \ref{fig:pie_testing} illustrates that in both countries a modest level of contact tracing was employed for most of the time.
    \item Governmental restrictions --- in specific \textit{'stay\_home\_restrictions'}, \textit{'school\_closing'} and \textit{'workplace\_closing'}, were used to build underlying contact networks and a varying exposure parameter. 
\end{itemize}

\subsubsection{Vaccination}\label{sss:vaccination}
We have available the total number of vaccines administered each day for each country. Therefore, we randomly select the corresponding number of nodes and classify them as vaccinated. When choosing the potential nodes to get vaccinated, we exclude deceased, hospitalized, symptomatic, quarantined, already vaccinated, and the nodes representing children (for each graph, we take a specific percentage of nodes representing children). We further use the parameters $v_{eff1}$ and $v_{eff2}$ to account for different impacts of the vaccine on the virus. When a vaccinated susceptible node comes into a contact with an infected neighbor, the probability that it gets exposed is reduced from $p_e$ to $(1-v_{eff1})p_e$ (in Figure \ref{fig:scheme} this effect is denoted with $p_e|v_{eff1}$). Additionally, if a vaccinated node is already infected, the chances that it develops symptoms are reduced from $p_{sy}$ to $(1-v_{eff2})p_{sy}$ (this means that the probability of being asymptomatic is increased accordingly), and in Figure~\ref{fig:scheme} this effect is denoted with $p_{sy}|v_{eff2}$ (and $1 - p_{sy}|v_{eff2}$).

\subsubsection{Testing and Contact Tracing}\label{sss:testing_tracing}
\begin{figure}[h]
    \centering
    \includegraphics[width=0.95\linewidth]{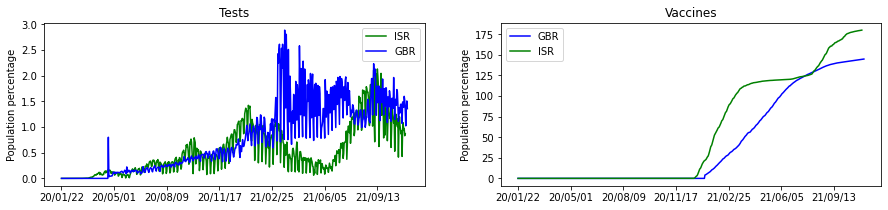}    \caption{Second group of predictors. Fraction of time (within a span of data) when each policy was in place.}
    \label{fig:pie_testing}
\end{figure}

When a testing policy is in place (available in the data along with the number of daily tests --- see Figures \ref{fig:test_and_vaccines} and \ref{fig:pie_testing}), there can be random testing or exclusive testing of symptomatic individuals. In case of random testing, we sample nodes that are not already quarantined, deceased, hospitalized, or recovered. All individuals with a positive test go to quarantine, so they move to either $Q_{Asy}$ or $Q_{Sy}$, depending on their current state. 

Nodes can also transition to quarantine through \textit{'contact\_tracing'} - when this policy is active, nodes that were in contact (i.e., share an edge) with a node that tested positive get automatically quarantined. This variable has 3 levels, contemplating: 1 --- no contact tracing, 2 --- limited contact tracing, or 3 --- extensive contact tracing (see Figure \ref{fig:pie_testing}). For levels 2 and 3, we consider a parameter $p_{ct}$ that controls the percentage of the contacts that gets traced. (Ideally, the value for the third level $p_{ct}(3)$ should be 1, but this is not realistic, and instead we take $0<<p_{ct}(2)<p_{ct}(3)<1$.) When this measure is active, states $Q_E$ and $Q_S$ of our model (Figure \ref{fig:scheme}) are feasible as well.

\subsubsection{Social Networks}\label{sss:social_networks}
Every day $t$ and for each region, we consider a network $G(t)$ representing the observed population. Each individual is a single node in the graph, and the edges between nodes correspond to contacts between people. The topology of $G(t)$ depends on the restriction policies that are in place at day $t$, hence we need to consider different underlying graphs in order to model the population for various restriction scenarios. Here we consider \textit{'stay\_home\_restrictions'} (a binary flag), \textit{'school\_closing'} (four levels of restrictions) and \textit{'workplace\_closing'} (four levels of restrictions), and similarly to \cite{networks}, we introduce a series of lattice and random graph corresponding to each combination of imposed measures:

\begin{itemize}
\item Regular Lattice $L_0$: each node $i$ is connected to 4 other physically close to $i$. This network is used to model the population when hard isolation is enforced, which means that all three flags are set to the highest level of restrictions. Theoretically, hard isolation should correspond to separate small connected components (corresponding to different households), but in practice, this representation is known to be more realistic as it takes into account a certain level of unavoidable interaction between individuals in different households \cite{networks}.
\item Small-World Lattice $L_{\alpha}$ \cite{rewired}: starting from a regular lattice with degree 4, each edge $(i,j)$ has a (small) probability $\alpha$ of being replaced with $(i,k)$ where $k$ is a random node in the graph. Typically here the rewiring probability $\alpha$ is chosen close to zero, and we use these networks to represent situations where the population is not in lock-down, but there are still strong limitations in place regarding gatherings and movement.
\item Random Graph $R_{\beta}$ (Erdos-Renyi network) \cite{ER}: each possible edge between two nodes is created with probability $\beta$. This network, even in the case when $\beta$ is chosen in such a way that the average degree is 4, provides a much larger level of connectivity than the rewired lattice $L_{\alpha}$ (unless $\alpha$ is close to 1), because the number of interactions between individuals that are physically far away from each other will be significantly larger. We use these networks when only mild restrictions (or no restrictions at all) are enforced.
\end{itemize}

\begin{figure}[h!]
    \centering
    \includegraphics[width=0.95\linewidth]{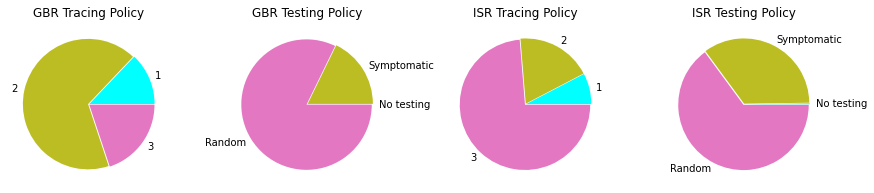}    
    \caption{Vectors indicating when each of 8 underlying graphs is applied, over a given time series.}
    \label{fig:graphs}
\end{figure}

By progressively combining the above networks, we create different levels of contact between the population. In order to do so, we start with a four-degree lattice graph $L_0$ and sequentially compose it with three other small lattice graphs and afterward with four random graphs as to grow the contact between nodes. In total, this yields 8 different networks built in the following way:
\begin{equation}
    \centering
\begin{split}
L_s &= L_0 \cup L_{p_l}\\
L_m &= L_s \cup L_{p_l}\\
L_l &= L_m \cup L_{p_l}\\
R_{xs} &= L_l \cup R_{p_{rxs}}\\
R_{s} &= R_{xs} \cup R_{p_{rs}}\\
R_{m} &= R_{s} \cup R_{p_{rm}}\\
R_{l} &= R_{m} \cup R_{p_{rl}}\\
\end{split}
\end{equation}
The fact that we build each network on top of the previous one ensures that we do not introduce new edges when transitioning from the denser graph to a more sparse one (when the restrictions get more strict). During the simulation, on each day a suitable network is chosen according to the restrictions in place as explained in Table \ref{tab:networks}. Abbreviations from the table stand for: WPCf: workspace closing flag; SCf: school closing flag; LDf: lockdown flag. Figure \ref{fig:graphs} shows the temporal positions of the graphs for each country. Note that we leave the network probabilities ($p_l, p_{rxs},p_{rs}, p_{rm}, p_{rl} $) as model parameters that should be later optimized. The reasoning is that they represent population behavior with respect to the measures and so should be tuned separately to each country.
\begin{table}[h!]
\centering
\caption{Underlying networks for different combinations of school and workplace closing levels and lockdown restrictions. A lower level entails less restrictions, increasing up to $7$, which represents the full lockdown situation.}\label{tab:networks}
\begin{tabular}{c|cccccccc}
  WPCf + SCf +LDf  &  $0$    &   $1$    & $2$      & $3$         &   $4$     &  $5$   & $6$     &  $7$     \\
  \hline
      Graph             & $R_l$ & $R_m $ &  $R_s$ &  $R_{xs}$ &  $L_l$ & $L_m$ & $L_s$ & $L_0$ \\ 
\end{tabular}
\end{table}

\subsubsection{Probability of exposure $p_e$}\label{sss:prob_exposure}
\begin{figure}[h]
    \centering
    \includegraphics[width=0.6\linewidth]{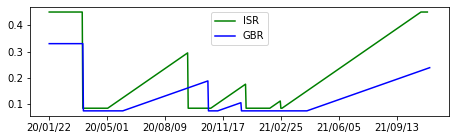}    \caption{Evolution of $p_e$ over time.}
    \label{fig:pevector}
\end{figure}

The probability of being exposed after the contact with an infected individual depends on the virus characteristics, but also on what kind of contact people have with each other. For instance, with the progression of the pandemic people have become more aware of the measures to be taken when in contact with others. We assume that this probability varies with the restrictions in place, and we model it as a function of three parameters ($p_{e_{max}},p_{e_{min}}$ and $t$) and a lockdown flag. When there are more severe restrictions, people are more careful in their contacts, so this value should be lower and vice-versa. We assume that there is maximum $(p_{e_{max}})$ and minimum $(p_{e_{min}})$ value that the parameter $p_e$ can take (in the extreme cases). Before the first lockdown occurs, $p_e$ keeps its maximum value ($p_{e_{max}}$), and as soon as the lockdown is imposed it drops to its minimum ($p_{e_{min}}$). After lifting the lockdown, $p_e$ starts linearly increasing so that after $t$ months it attains its maximum value again (unless another lockdown is imposed). This behaviour is enforced in the same way any time a new lockdown occurs. The estimated $p_e$ vectors for each country are presented in Figure \ref{fig:pevector}.

\subsection{Decomposition and Complete Model}\label{ss:decomposition}

Instead of creating a single large graph to represent a country's population, we consider several smaller graphs corresponding to separate cities or densely populated regions that have fewer connections with the rest of the population. While this has the primary goal of enforcing a more realistic simulation setting, it provides an additional advantage in terms of computational costs.
When the size $N$ of the considered population is large, working with a network of $N$ nodes becomes too expensive, but by splitting it into $m$ subgraphs of sizes $N_i<<N$ for $i=1,...,m$, we reduce this cost considerably and allow for parallel processing. 
 At each time step, for each separate network, we simulate the progression in the spread of the infection according to the model described in the previous subsections. Then, we model the interactions among the networks following a standard (not network-based) compartmental approach. That is, once the states of all nodes have been updated, we take into account the exposures that may arise from contacts between people in different networks. For every network indexed $i=1,...,m$ we define the additional set of exposed people as
\begin{equation}\label{mixing}
 \hat E_i(t) = \sum_{j\neq i} w_{ij}(t)\frac{p_e}{N}I_i(t)S_j(t)
\end{equation}
where $p_e$ is the probability of an individual becoming exposed after being in direct contact with an infected person and $w_{ij}(t)\in[0,1]$ is a parameter that regulates how likely it is for individuals in the region $i$ to enter in contact with individuals of the region $j$. We can see that the above equation is analogous to the one used in the classical SIR model \eqref{eq:SIR}: the only significant difference is that here we assume that individuals become exposed before transitioning into the class of infected people, and that we weight the number of expected contacts by multiplying with the coefficient $w_{ij}$. The choice of the coefficient $w_{ij}(t)$ will depend on the restrictions that are enforced at the considered time $t$.\\ 
This decomposition allows for the application of an approximated network-based approach to the case of populations of entire countries. A further advantage is that the decomposition can be done in networks of equal size or in networks representing the country's administrative regions --- an appealing setting if different restrictions are enforced in different areas. 

 Finally, in order to account for people entering the country while infected, we introduce a parameter $p_{int}$, active only when there are no international travel restrictions, that represents the probability of introducing a new exposed case in each region. This parameter is set to a low value $p_{int}=0.01$, since higher values would make the model deviate too far from the closed system setting, and setting it to zero would reduce a realistic representation. With additional information on international travel, one could potentially improve this estimate.

\section{Numerical Simulations}
\label{sec:NumSim}

\subsection{General setting}

As mentioned before, the population of a country is simulated with an underlying sparse graph. In order to execute simulations on a user-level PC, we cannot work with the exact population size but reduce it by a factor of 100. Further, to compensate for the fact that population is not actually homogeneous but consists of several connected components (cities or regions) that have only sporadic contacts, we compose the population graph $G$ as a union of disjoint subgraphs $G_1,...,G_m$. Each subgraph $G_i$ consists of $n_i$ nodes, where $25000<n_i<3500$. Since the regions obtained in this way are disjoint, we model the "inter-regional contact" using a differential compartmental model as described in Section~\ref{ss:decomposition}. This will introduce additional exposed cases, proportional to the number of currently infected individuals in the neighboring regions, unless internal traveling restrictions are in place. For our simulations we assume that the parameter $w_{ij}$ of equation \eqref{mixing} is the same for every index $i$ and $j$ and equal to $r_{mix}$. Since this parameter models the number of contacts among individuals in different cities or regions, it is reasonable to keep its value small. We assume in our simulations that $r_{max}\leq0.1$. It is important to note that this approach to modeling the population is also favorable in terms of computational costs --- working on a single connected graph of the same size as $G$ is infeasible on user-level machines. 

Since the data shows that the first cases appear after 30 days of recording, we introduce one exposed node per region at day 30 of simulation. As an approximation for the number of people that should be excluded from vaccination, we take the percentage of children in each country ($18\%$ for GBR and $27\%$ for ISR)\footnote{For more accurate modeling, these percentages could be augmented with people that will not get vaccinated due to due to some other reasons (medical, religious,...)}.

Simulations are run over more than 20 months, starting on January 2020, until November 2021. Every simulation with each set of parameter values is executed ten times; this, together with randomness in graph sizes and connections, has showed to be enough to give a good generalization of a scenario. 

\subsection{Parameter optimization}\label{subsec:crossvalidation}

We distinguish between parameters related to the virus biology and parameters related to the population's particular characteristics or behavior. We consider the first as country-independent and assume that their values should be in a similar range as proposed in the literature. On the other hand, parameters related to the population should be country-dependent (thus, we allow different values for ISR and GBR during the optimization) and cannot be found in the literature, as they are specific to our model and often have very different meanings. Both groups are listed in Tables \ref{tab:val_literatures_independent} and \ref{tab:val_literatures_dependent}, with (available) literature ranges and our final estimates. 

\begin{table}[h!]
\caption{Country-independent parameters with literature ranges suggested in the literature. The probability of infection is inverse to the number of incubation days (13 days), the probability of recovering is inverse to the average length of recovery (10 days), probability of leaving the quarantine is inverse to the length of quarantine (14 days). Vaccine efficacy averages around 0.9. }\label{tab:val_literatures_independent}
\begin{tabular}{c | c | c | c}
Parameter     & Literature range   &  Source                  & Our estimate \\ \cline{1-4}
$(v_{eff1},v_{eff2} )$     & $0.8 - 0.95 $      &  \cite{schools,vaccines} &        $(0.95,0.7) $     \\
$p_i$         & $3^{-1} - 20^{-1}$ &  \cite{LiSubstantial,LauerIncubation,ReadNovel,EikenberryInfectious,TangUpdated,ZhouClinical} & $0.08$ \\
$p_{sy}$     & $0.13 - 0.65$      &  \cite{TangUpdated,LiSubstantial,FergusonReport,EikenberryInfectious,MoriartyPublic}    & $0.5$ \\
$p_r$         & $3^{-1} - 30^{-1}$ &  \cite{ZhouClinical,TangUpdated,EikenberryInfectious} & $1/29$ \\
$p_{syh}$     & $0.05 - 0.15$      &  \cite{EikenberryInfectious} & $0.006$\\
$p_{hd}$      & $10^{-5} - 10^{-1}$& 
\cite{TangUpdated,EikenberryInfectious,FergusonReport,ZhouClinical,VerityEstimates} & $0.03$ \\
$p_{hr}$      &$3^{-1} - 30^{-1}$ &  \cite{ZhouClinical,TangUpdated,EikenberryInfectious} & $1/11$
\end{tabular}
\end{table}

\begin{table}[h!]
\caption{Country dependent parameters, with initial and final estimates for GBR and ISR. }\label{tab:val_literatures_dependent}
\begin{tabular}{c | c | c | c}
Parameter     & Initial range   & GBR estimate& ISR estimate \\ \cline{1-4}
$(p_{e_{min}},p_{e_{max}})$ & $(0.01 - 0.2, 0.2 - 0.5)$ &$(0.075,0.33)$ & $(0.085,0.45)$\\
$t$ & $3 - 18$ & $11 $& $8$ \\
$(p_{ct}(2),p_{ct}(3))$ & $(0.3 - 0.7, 0.7 - 1)$ & $(0.65,0.8)$&$(0.65,0.8)$\\
$p_l$ & $0.004 - 0.008$ & $0.006$ & $0.004$\\
$(p_{rxs},p_{rs},p_{rm},p_{rl})$ & $0.5 - 4.5$ &$(1.5,0.8,1.5,0.8)$ &$(0.9,1.5,1,0.8)$\\
$r_{mix}$ & $0.05 - 0.15$ &$0.065$ &$0.065$\\
\end{tabular}
\end{table}

Regarding the first group of parameters, numerous papers proposed the values for the compartmental models concerning COVID-19; however, most of them focus only on a single country or a region or were estimated early in the pandemic, hence we should be careful to accept them. Still, we might consider that parameter values obtained in different studies give a reliable interval range as a starting point in optimization. Therefore, we take values suggested across the literature to narrow down the grid search space when fitting the values for ISR and GBR. 

An additional remark is due regarding vaccination. The information about vaccination present in our data only refers to the total number of doses given per day, without detailing how many doses each individual received, which leads us to consider that one dose is enough for a person to be vaccinated. However, this means that we should allow for lower values of $v_{eff1}$ than the ones found in the literature.

\subsubsection{Parameter tuning}\label{sss:parameter_tuning}

We use two metrics in the optimization process: sMAPE \cite{article:smape} and Pearson correlation \cite{article:pearson}. Both functions are altered to fit our data-series structure, and performed over smoothed curves for the number of hospitalized and deceased cases per day. As mentioned in Section \ref{subsec:Datasets}, we choose these two attributes as more reliable, but we also include an additional exclusion criterion that considers the cumulative number of cases --- if the simulation terminates with a too high or too low number of cases, it is automatically excluded from further consideration. That is, to ensure that the predictions are reasonable, we exclude combinations where the percentage of exposed nodes is above a given threshold and the percentage of symptomatic ones is below another. This exclusion is activated if any of 10 runs violates the constraints, and it additionally ensures that our model does not overfit.

Our optimization process was split into two stages. In the first stage, we create random combinations of feasible parameter values and progressively reduce the search space based on the results. After this initial step, we start with the combination that scored the best results, and further optimize each parameter individually, by turns, within a reduced range. In the initial phases of the process, the above metrics in combination with qualitative evaluation were used to select the optimal value, as the results were visually diverse. In later stages, only the objective metrics were taken into account to refine the fitting.

\subsection{Final Predictions}

\begin{figure}[h]
    \centering
    \includegraphics[width=0.95\linewidth]{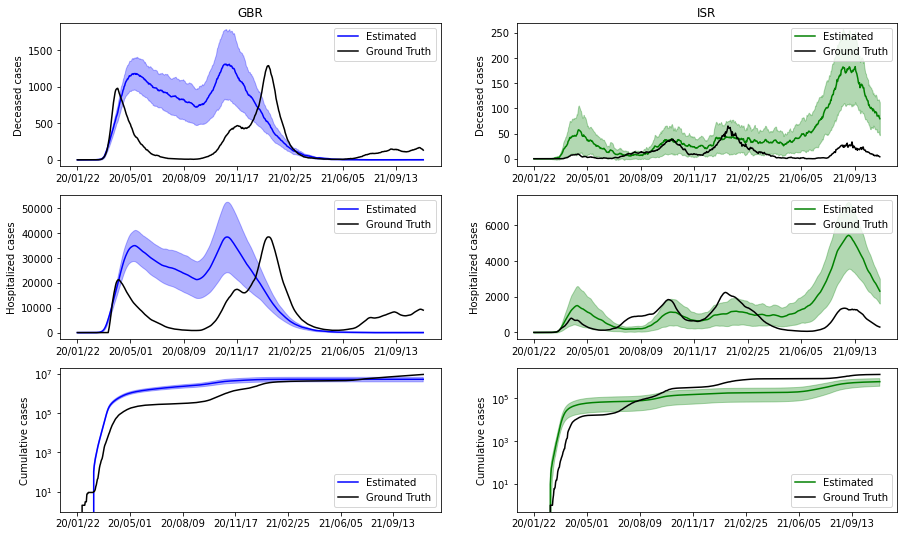} 
    \caption{Final estimated solution. Left column: GBR; right column: ISR. Ground truth data is presented by a solid black line, while estimated numbers are presented in blue for GBR and in green for ISR --- solid lines indicate mean over 100 simulations, and shaded area shows standard deviation.}
    \label{fig:estimated_solution}
\end{figure}

Simulations with the optimal set of parameter values were run 100 times, and the results for both countries are presented in Figure \ref{fig:estimated_solution}. The mean over predictions is presented in solid lines, while the shaded area corresponds to standard deviation (divided by two). An important remark is that both the mean and standard deviation reached a stable solution after only 10 trials, without noticeable changes in the subsequent 90.

Even if we do not observe a perfect fit for the curves, it is noticeable that the main waves are farily well predicted in terms of time and strength --- the main miss fit is at the final stage. For that period, our simulations are not able to predict the third wave in GBR and produce an overestimation of cases for ISR. A probable explanation lies on the long term fitting that we are taking into account (almost two years). A more realistic model could consider possible loss of immunity for recovered and vaccinated nodes, or even natural demographic changes. This is a possible line of research to be addressed in future work.

It is pertinent to compare the estimated values for country dependent parameters in Table~\ref{tab:val_literatures_dependent}. A larger $p_{e_{max}}$ value for ISR suggests more contacts in the absence of restrictions for this country. In addition, a lower value of $t$ suggests that after a lockdown ISR is faster to return to the usual state of contact levels. Interestingly, the percentage of contact tracing, even if left free to be individually tuned, was found to take similar values, which suggests levels 2 and 3 have the same meaning for both countries. Finally, looking at the network parameters we see that $p_l$ takes a lower value for ISR, possibly entailing that the restrictions closer to lockdown are enforced more strictly there. For the remaining topologies, given the considered ranges, we find that the estimated parameters are in general similar, especially for the most relaxed networks (i.e. closer to a non restricted scenario. This is reasonable, since $p_{e_{max}}$ should already be expressing any differences regarding this situation.

\section{Impact of Different Measures}
\label{sec:Scenarios}
In this section we use the proposed model to study different hypothetical scenarios. This can be useful in order to understand the impact of different policies and restrictions and therefore in providing helpful information that may guide the decision process. Unless otherwise stated, the parameters used in the simulations are the ones obtained in the previous section. Moreover, since the goal is to evaluate the behaviour of the model when different measures are in place, we compare the scenarios with the original setting when pertinent.

\subsection{Restriction measures}

\begin{figure}[h]
    \centering
    \includegraphics[width=0.95\linewidth]{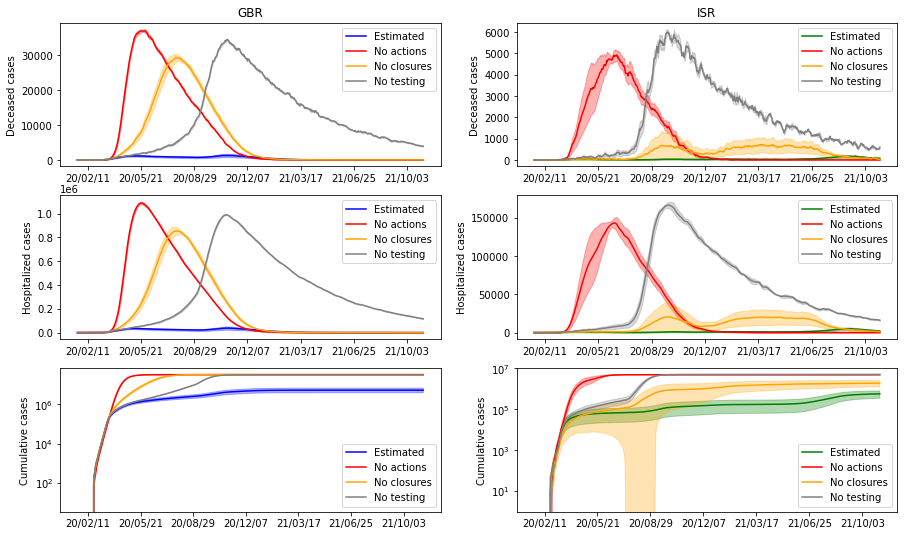} 
    \caption{\textit{Scenarios  (1,2,3)}. Comparison between three different relaxations on containment measures and the true ones. \textit{Scenario 1}, labeled 'No actions', considers the evolution of the pandemic without any government intervention. \textit{Scenario 2}, labeled 'No closures', takes into account testing and contact tracing, but no further restrictions (e.g. lockdown or workplace closing).  \textit{Scenario 3}, labeled 'No testing or tracing', is a complement setting of \textit{Scenario 2}, with restrictions on movement but no contact tracing or testing in place. Vaccination campaign proceeds as in the data for the latter two settings. We present cumulative cases, hospitalized and deceased for both GBR (on the left) and ISR (on the right).}
    \label{fig:scenarios1_71_62}
\end{figure}

In \textit{Scenario 1}, we consider the case where no action at all is taken to try to contain the spread of the disease. We assume here that no stay-at-home policy is mandated by the government and that schools and workplaces are open without any restrictions. Moreover, we assume that no tests are performed to try and detect infected individuals and that no contact tracing strategy is in place (in particular this implies that there is no mandatory quarantine for anybody). We also assume that the vaccination campaign never starts.\\
In \textit{Scenario 2} we consider the case where no generalized restrictions are mandated by the government, but all the other policies such as testing, contact tracing (with quarantine) and vaccinations are still in place and are the same as in the data.\\
In \textit{Scenario 3} we consider the case where stay-at-home and work/study-from home policies are enforced according to data, but neither testing nor contact tracing policies are employed, hence no infected individual gets quarantined, regardless of their state. We also assume that vaccination campaign proceeds as in the data.
In Figure \ref{fig:scenarios1_71_62} we plot the predictions for these three scenarios. As we can see, they all lead to a spread of the disease that is much larger than the one given by the actual policies, suggesting that, as expected, none of the three approaches represented by the described scenarios are viable in practice. In particular, we can notice that limitations on daily activities such as working and study from home policies are essential to contain the spread of the virus, as well as testing procedure and planned quarantine time for infected individual. Lifting any of these containment measures seems to lead to the statistics comparable with the case where no action was taken. In Figure \ref{fig:scenarios1_71_62}, we see that in all three scenarios we have a single mode, i.e. the virus quickly reaches the point of saturation, except in the scenario of no imposed closures in ISR. This last case is in accordance with the scenario that we will consider next, and indicates that the stay home policy was enforced more strictly in ISR than in GBR. From the same results we can also conclude that, while both restrictions and testing are essential, neither of these policies alone is enough to sufficiently limit the spread of the disease.\\

\begin{figure}[h]
    \centering
    \includegraphics[width=0.95\linewidth]{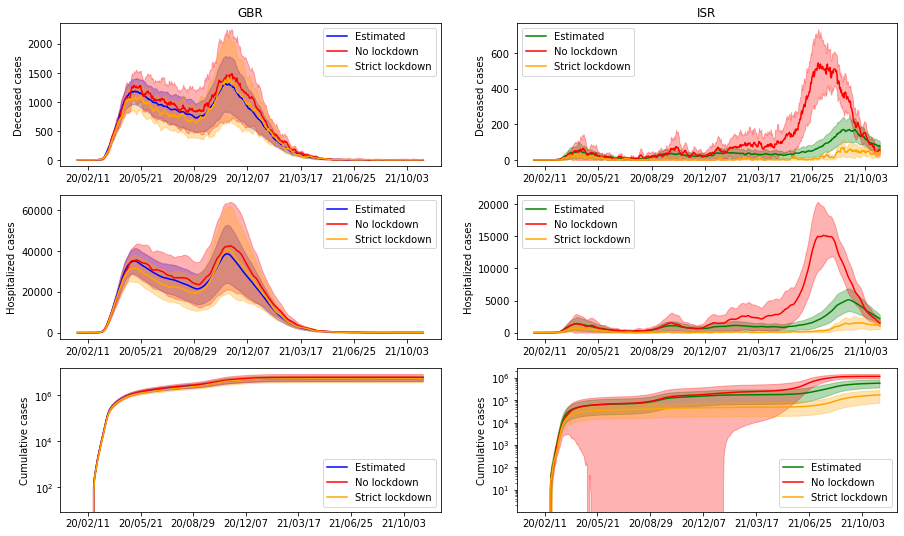} 
     \caption{\textit{Scenarios  (4,5)}. Comparison between milder relaxations on restriction policies, confronted with the real ones. \textit{Scenario 4}, labeled 'No lockdown', considers closing of schools and workplaces, but absence of stay-at-home policy. All other measures remain unaltered. \textit{Scenario 5}, labeled 'Strict lockdown', considers that whenever a lockdown is in place, schools and workplaces are also closed, but than no limitations on schools and work attendance are imposed when stay-at-home policies are not enforced. We present cumulative cases, hospitalized and deceased for both GBR (on the left) and ISR (on the right). }
    \label{fig:scenarios7_8}
\end{figure}

We now study milder relaxations of the restrictions, with respect to the data. The obtained results are plotted in Figure \ref{fig:scenarios7_8}.
In \textit{Scenario 4}, we consider the case where no stay-at-home policy is enforced but closing of schools and workplaces (as well as all other measures such as quarantine and vaccinations) are mandated as in the original data. In \textit{Scenario 5}, we assume that whenever a lockdown is in place, schools and workplaces are also closed, but than no limitations on schools and work attendance are imposed when stay-at-home policies are not enforced.\\

As we can see from Figure \ref{fig:scenarios7_8}, these two choices lead to a number of cases that is similar to our estimate for the original data. The results for \textit{Scenario 5} suggest that limiting the physical attendance of schools and workplaces only for the period of times where stay-at-home policies are also enforced may be a viable option. This is particularly interesting because it could result in a significantly shorter amount of time during which students and employees are forced to work from home, without leading to major changes in the spread of the virus. At the same time, we see that as long as schools and workplaces are subject to restrictions, stay-at-home policies seem to be inessential: for the two countries that we took in examination this seems to make a significant difference only for the end of the considered time-frame and only for Israel. The reason why GBR seems to be less affected by the changes in\textit{ Scenario 4} and \textit{Scenario 5} with respect to the original data is that, for the vast majority of the considered days, stay-at-home and work/study-from-home policies were mandated and lifted at the same time. In ISR, on the other hand, several different combinations of restrictions where active in different days during the considered time interval.

\subsection{Vaccines}

In \textit{Scenario 6} we study how the spread of the virus among the population is influenced by the vaccine efficacy. We assume that the vaccination campaign proceeds in the two countries as in the considered data, but we consider the case where the efficacy of the vaccine is given by a factor $k$ times the efficacy found in the previous section, for $k\in\{0,0.5,1.2,1.5\}$. In \textit{Scenario 7}  we investigate the consequences of different vaccination rates on the evolution of the contagion. We assume that the vaccine efficacy is the same as in the previous section and that the initial day of the vaccination campaigns in the two countries is the same as in the data, but we vary the amount of administered vaccines. Specifically we consider the case where the number of vaccines distributed to the population per day is $k$ times the real number of daily vaccines with $k\in\{0,0.5,1.5,2\}$. We remark that in both \textit{Scenario 6} and \textit{Scenario 7}, a factor $k=0$ naturally corresponds to the case where no vaccines are administered at all.\\

\begin{figure}[h]
    \centering
    \includegraphics[width=0.95\linewidth]{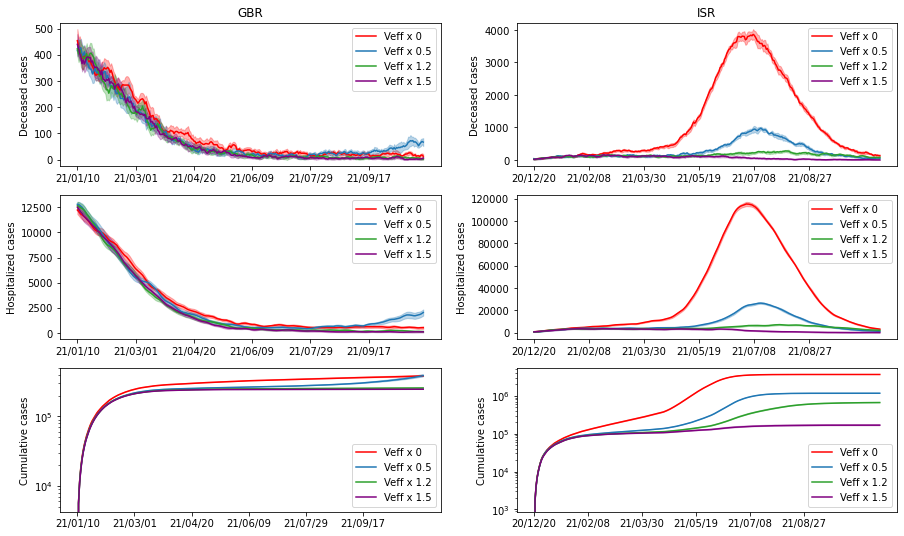} 
    \caption{ \textit{Scenario 6}. Comparison of simulations with different vaccine efficacy, where $k$ is a factor multiplied by the estimated value found in our estimate. All other settings remain the same. We present cumulative cases, hospitalized and deceased for both GBR (on the left) and ISR (on the right). Note that the simulation starts only on the first day of the vaccination campaign for each country. }
    \label{fig:scenario_veff}
\end{figure}

\begin{figure}[h]
    \centering
    \includegraphics[width=0.95\linewidth]{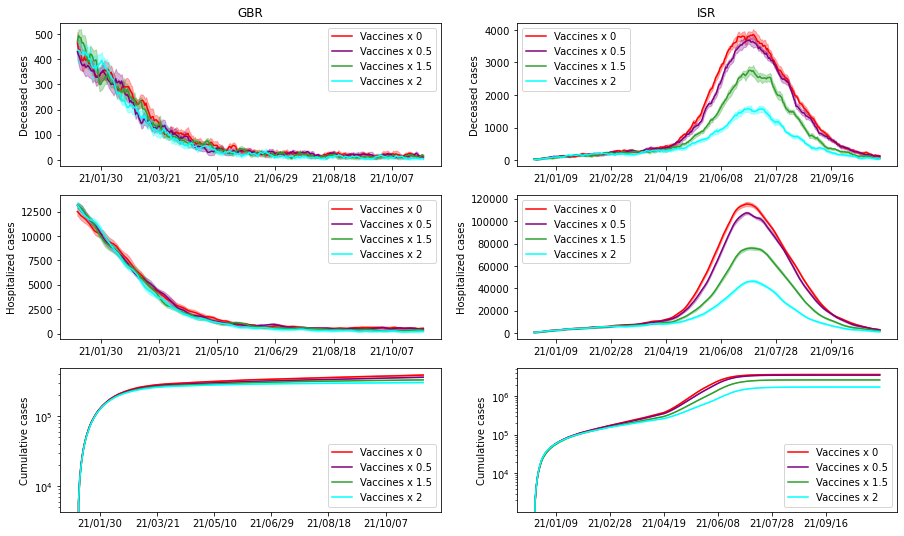} 
     \caption{\textit{Scenario 7}. Comparison of simulations with different numbers of vaccines administered. Parameter $k$ is factor multiplying the average daily number of vaccines according to the available data for each country. Vaccine efficacy is the one previously estimated, as well as the remaining parameters. We present cumulative cases, hospitalized and deceased for both GBR (on the left) and ISR (on the right). Note that the simulation starts only on the first day of the vaccination campaign for each country.}
    \label{fig:scenario_vnum}
\end{figure}

In Figures \ref{fig:scenario_veff} (\textit{Scenario 6}) and \ref{fig:scenario_vnum} (\textit{Scenario 7}), we plot the obtained results. In order to have a fair picture of how vaccines affect the outcome, we need to have the same starting point. We accomplish this by starting simulations at the initial day of the campaign (day $333$ for ISR and day $354$ for GBR), with the same initial state. That is, we start with a graph whose nodes are assigned to corresponding states according to the average of our initial estimate for the considered day. For completeness, we include both countries, but since in the considered timeframe the vaccination campaign of GBR started relatively late, we can notice that different vaccination policies and efficacy have no significant impact on the evolution on the spread there. Nonetheless, we point out that for vaccine efficacy reduced by a factor of 2 there is a start of a second peak, which is also present in the real data. This may originate from our simplified vaccine model, where we consider one person fully vaccinated after a single dose. From Figures \ref{fig:scenario_veff} and \ref{fig:scenario_vnum}, we can see that, as expected, not vaccinating the population leads to a significantly higher number of hospitalized and deceased people, compared to any other vaccination campaign. Moreover, comparing the two Figures, we can see that the vaccine efficacy has larger impact on the number of cases and hospitalized people when compared to the vaccination rate. This suggests that, while having promptly available vaccines and proceeding rapidly in the distribution to the population certainly helps to reduce the number of cases and in particular the number of people seriously affected by the disease, the quality of the vaccines plays a mayor role. \\

\subsection{Testing and contact tracing}

To study the influence of the testing and contact tracing procedures on the spread of the disease, we consider three different scenarios:
\begin{itemize}
    \item varying the number of tests;
    \item varying the initial date of the testing campaign;
    \item varying the accuracy of the contact tracing policy.
\end{itemize} 
In general, the more tests are carried out, among both people that manifest symptoms and the general population, the more infected individuals are detected and therefore sent to quarantine, which should significantly limit the spread of the virus. A similar remark can be done for contact tracing policies --- the more accurate the contact tracing procedure (so, the higher the amount of direct contacts that is quarantined), the smaller is a risk of potentially infected individuals spreading the virus. \\

In \textit{Scenario 8} we assume that the contact tracing and testing policies are the same as in the data, but that the number of tests is different. Specifically we consider the case where the number of tests performed at each day of the simulation is $k$ times the real number of daily tests with $k\in\{0.5,1.5,2\}$. In \textit{Scenario 9}, we study how the starting date of the testing procedure influences the spread of the virus. We assume that the testing campaign begins at day $t_0$ and we consider several values of $t_0$ between 30 and 180. To simplify the comparison we assume that at each day $t$ after $t_0$ a fixed number of tests (equal to the daily average of test performed by the country over the considered timeframe) is performed over the population.

\begin{figure}[h]
    \centering
    \includegraphics[width=0.95\linewidth]{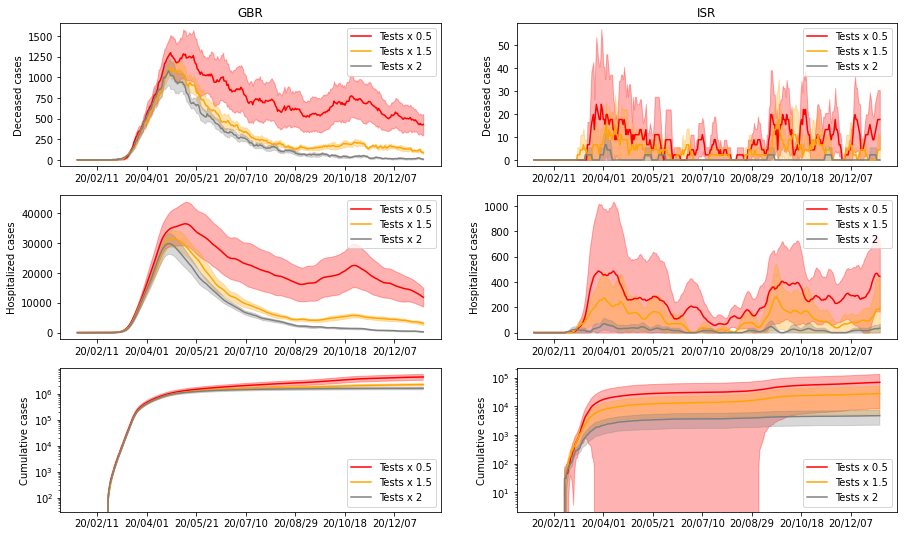} 
      \caption{\textit{Scenario 8}. Comparison of simulations with different number of tests performed, without alterations on tracing and testing policies. The factor for each case is applied to the real number of daily tests. We present cumulative cases, hospitalized and deceased for both GBR (on the left) and ISR (on the right).}
    \label{fig:scenario_num_tests}
\end{figure}

\begin{figure}[h]
    \centering
    \includegraphics[width=0.95\linewidth]{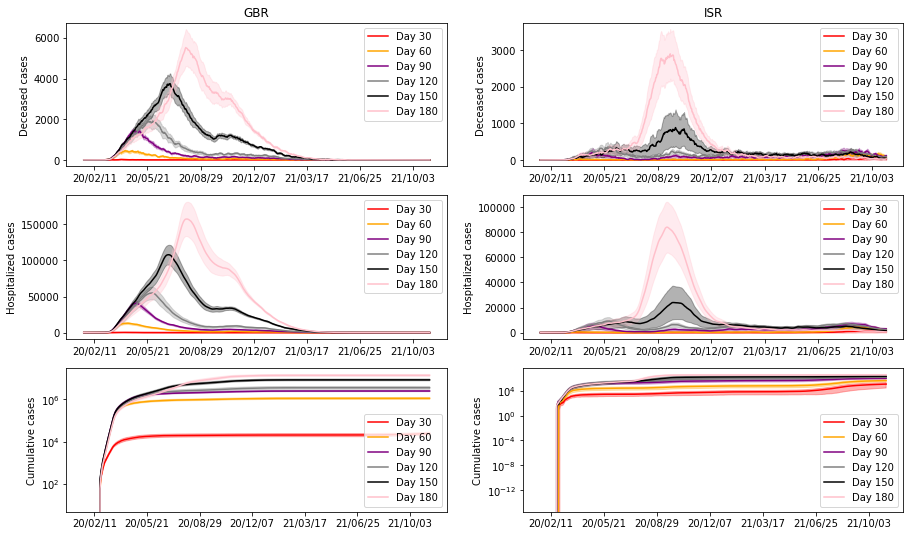} 
    \caption{\textit{Scenario 9}. Comparison of simulations with different starting day of test campaign, without alterations on tracing and testing policies. We present cumulative cases, hospitalized and deceased for both GBR (on the left) and ISR (on the right).}
    \label{fig:scenario_test_start}
\end{figure}

Results for \textit{Scenario 8} can be found in Figure~\ref{fig:scenario_num_tests} and for \textit{Scenario 9} in Figure \ref{fig:scenario_test_start}. As expected, the increase in daily tests leads to a decrease in number of cases. We further note that the case of no tests (i.e. a factor of $0$) can be found in Figure~\ref{fig:scenarios1_71_62} under the label 'No testing or tracing' and follows the same pattern. When it comes to the starting day, the differences are visible from the beginning, and the simulations with a late start of testing exhibit extremely large peaks for deaths and hospitalizations. In the final days, all the simulations go down towards zero, so in Figure \ref{fig:scenario_test_start} we show only the first 300 days, for sake of clarity. The main conclusion to be drawn is that a promptly start of testing campaigns may be of great help in significantly limiting and potentially stopping the spread of the virus. It is also pertinent to compare the two scenarios --- the variation achieved in Figure~\ref{fig:scenario_num_tests} for different factors is almost negligible, when compared to the one in Figure \ref{fig:scenario_test_start}. This suggests that the latter is more important to control the number of cases. \\

\subsection{Cases-driven lockdown}
\begin{figure}[h]
    \centering
    \includegraphics[width=0.95\linewidth]{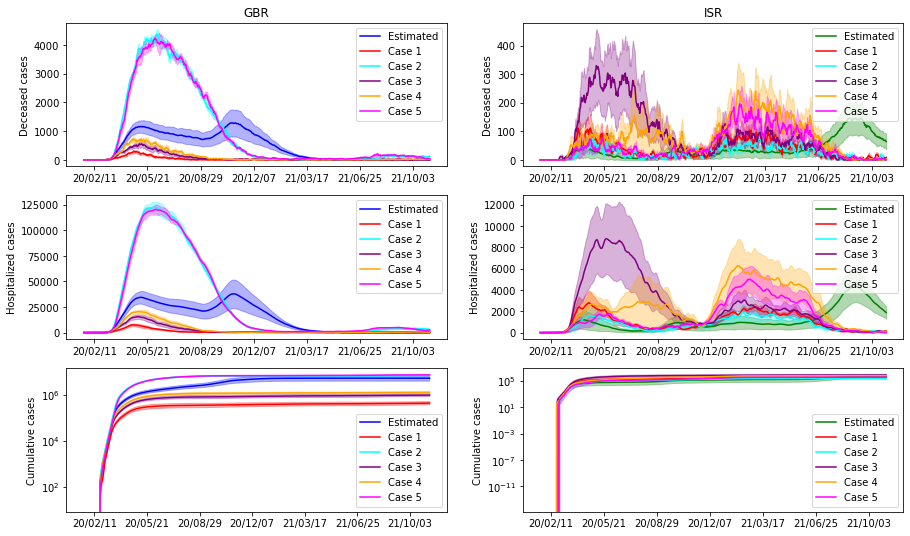} 
     \caption{\textit{Scenario 10}. Simulations with cases-imposed lockdown, for different combinations of parameters $h_{lock}$ and $t_{lock}$ (see Table~\ref{tab:open_close}).We present cumulative cases, hospitalized and deceased for both GBR (on the left) and ISR (on the right).}
    \label{fig:scenario_openclose}
\end{figure}

Finally, we test \textit{Scenario 10}, where a lockdown is activated when a given number of hospitalized cases is reached. This threshold $h_{lock}$ is set as a percentage of the total population and we test different values for it. The lockdown is then kept until the cases stay below the same threshold for a given period $t_{lock}$, also tested with different values. We show 5 different combinations of these 2 parameters, described in Table~\ref{tab:open_close}. Note that the threshold value was generally chosen to be smaller for GBR, since a value of $h_{lock}$ larger than $7.5^{-5}$ leads to an uncontrollable explosion in cases and so does not offer much insight. In Figure \ref{fig:scenario_openclose}, we see that cases 2 and 5 yield a high number of deaths for GBR, while the remaining options perform better than the original scenario. Another interesting insight is provided in Table~\ref{tab:open_close_days} showing the average number of days with imposed lockdown. We note that, for GBR, it is possible to reduce the number of causalities while enforcing lockdown restrictions of an inferior duration of what Figure~\ref{fig:scenario_openclose} might suggest. For ISR, we can notice two distinctive peaks for any considered scenario, which explains the longer lockdown duration in Table~\ref{tab:open_close_days}. Unlike GBR, we do not achieve an improvement for any of the proposed settings\footnote{We note a further reduction in the threshold value is not feasible in our experiments due to the simulated population size. This is an interesting line for future research.} and so, it would be relevant to consider a broader range of scenarios in order to potentially reach an effective strategy for both countries.

\begin{table}[h!]
    \centering
\caption{Parameters for different cases of \textit{Scenario 9}. }\label{tab:open_close}
\begin{tabular}{c | c | c | c}
Case     & $t_{lock}$   & $h_{lock}$ (GBR) &  $h_{lock}$ (ISR) \\ \cline{1-4}
1        & $7$  & $5\times 10^{-6}$     &  $5\times 10^{-5}$   \\
2        & $7$  & $ 10^{-5}$     &  $ 10^{-4}$   \\
3        & $14$ & $10^{-5}$     &  $10^{-4}$   \\
4        & $7$  & $7.5\times 10^{-5}$   &  $7.5\times 10^{-5}$ \\
5        & $14$ & $7.5\times 10^{-5}$   &  $7.5\times 10^{-5}$ \\
\end{tabular}
\end{table}

\begin{table}[h!]
    \centering
\caption{Average number of days with imposed quarantine for each case in \textit{Scenario 9}. }\label{tab:open_close_days}
\begin{tabular}{c | ccccc}
Case     &  $1$     &   $2$   & $3$    & $4$    & $5$  \\ \cline{1-6}
GBR      & $336$    & $315$   & $333$  & $336$  & $307$   \\
ISR      & $423$    & $358 $  & $427 $ & $428 $ & $409 $    \\
\end{tabular}
\end{table}

\section{Discussion}
\label{sec:Discussion}

In this paper, we proposed an extended SIR model that blends a standard compartmental (differential) model and a graph-based approach, fitted over two countries, in order to ensure better generalization properties. Estimated solutions show good behavior, correctly predicting positions and magnitudes of the peaks, except for the very last months of the time series. In order to increase accuracy of the final phase, additional assumptions should be considered, including loss of immunity induced by time, and perhaps booster doses of the vaccine --- we leave this as a direction for further research. 

Furthermore, we have investigated the effects of different containment measures w.r.t. the spread and the impact of the virus, by simulating alternative hypothetical scenarios with diverse combinations of the input parameters. We have found that testing plays a significant role in containing the outbreak, suggesting that an early start and frequent occurrence is fundamental. As expected, strategically imposing different levels of closure results in a decreasing (or increasing) number of cases. However, somewhat surprisingly, stay-at-home restrictions as an independent measure (i.e. without closing of schools and workplaces), do not show a great impact -- particularly for GBR, where the simulated scenario hardly differs from the original estimate.

Another interesting scenario is imposing (and lifting) the lockdown based on the number of hospitalized people at the moment. This strategy showed favorable results for GBR. For ISR, the considered scenarios were probably not diverse enough to achieve improvement in the number of cases. Besides a broader range of parameters, other possibilities of restrictions could be investigated in the future. 

The influence of the vaccination campaign shows clear and predictable patterns in the case of ISR --- if either the efficacy or the number of administered vaccines is increased, the number of cases decreases, and vice versa. For GBR, on the other hand, we observe almost negligible variations, which may be explained by their lower pace of vaccination and overall percentage of vaccinated people in the time-frame that we considered.

\section*{Acknowledgements}
This work was motivated by the participation in the ECMI 2021 Student Competition and we would like to thank the European Consortium for Mathematics in Industry (ECMI).

\section*{Funding}
This work has received funding from the European Union’s Horizon
2020 research and innovation program under the Marie Sklodowska-Curie
grant agreement No 812912.

\section*{Availability of data and materials}
The datasets generated and/or analysed during the current study are available in the Covid-19 Data Hub repository, https://covid19datahub.io/index.html \cite{DataHub}.\\

\section*{Authors' contributions}
All authors evenly contributed to the whole work. All authors read and approved the final manuscript.

\bibliographystyle{vancouver} 
\bibliography{bmc_article}

\end{document}